*Original Article*

# IoT-Based Solution for Paraplegic Sufferer to Send Signals to Physician via Internet

L. Srinivasan[1], D. Selvaraj[2], D. Dhinakaran[3], T. P. Anish[4]

*[1]Department of Computer Science and Engineering, Dr. N.G.P. Institute of Technology, Coimbatore, India*
*[2]Department of Electronics and Communication Engineering, Panimalar Engineering College, Chennai, India*
*[3]Department of Information Technology, Velammal Institute of Technology, Chennai, India*
*[4]Department of Computer Science and Engineering, R.M.K College of Engineering and Technology, Chennai, India*

*[1]Corresponding Author: srinivasanl.1982@gmail.com*

*Abstract - We come across hospitals and non-profit organizations that care for people with paralysis who have experienced all or portion of their physique being incapacitated by the paralyzing attack. Due to a lack of motor coordination by their mind, these persons are typically unable to communicate their requirements because they can speak clearly or use sign language. In such a case, we suggest a system that enables a disabled person to move any area of his body capable of moving to broadcast a text on the LCD. This method also addresses the circumstance in which the patient cannot be attended to in person and instead sends an SMS message using GSM. By detecting the user part's tilt direction, our suggested system operates. As a result, patients can communicate with physicians, therapists, or their loved ones at home or work over the web. Case-specific data, such as heart rate, must be continuously reported in health centers. The suggested method tracks the body of the case's pulse rate and other comparable data. For instance, photoplethysmography is used to assess heart rate. The decoded periodic data is transmitted continually via a Microcontroller coupled to a transmitting module. The croaker's cabin contains a receiver device that obtains and deciphers data as well as constantly exhibits it on Graphical interfaces viewable on the laptop. As a result, the croaker can monitor and handle multiple situations at once. The program also allows us to check the data collected. If any implied anomalies or changes in a case's status, a burglar alarm linked to the system will provide an audible alert message that a specific room's case needs immediate attention. The GSM modem attached to the device also transmits a signal to each of the croakers within this unit with the room number of the instance, which requires prompt attention in the event that the croaker isn't in his chamber. To solve this problem, we created a technique that enables such individuals to communicate with elementary motions. This gadget may be made to fit within a person's clothing or be put on their finger.*

## 1. Introduction

When a section or majority of the body is paralyzed, it loses its capacity to move and occasionally even its ability to feel. Spinal cord damage, stroke (severe or lateral), multiple sclerosis, etc., are common causes of paralysis. In addition, the nervous system can be injured or afflicted with disease(s), disrupting the nerve impulses transmitted to the limbs and causing paralysis [1,18]. This could lead to the following circumstances, including total lack of mobility or paralysis in any one arm, complete loss of motion or paralysis in both legs, or entire loss of mobility or paralysis across both arms solely on a single side of the body [2-4].

Today, paralyzed people are either left alone or observed by a nurse. Unfortunately, the Care Taker frequently leaves these sufferers alone without giving much thought to their fundamental needs. The proposed work aims to create wearable technology enabling patients to connect with their carers and monitor their health instantaneously. The IOT-based paralyzed patient health care system is an application created to assist the patient in communicating with doctors, nurses, or family members at home or working over the Internet. To accomplish this capability, the system uses electronics centered on a microprocessor. It uses a reception plus broadcaster circuitry and a hand gesture detection circuit [43]. The hand gesture circuit utilizes a gyroscope and accelerometer to identify arm movements and sends this data via wireless over RF to the transceiver. The transmitter system is intended to take these instructions, analyze them, and show the results on the LCD screen while transmitting the information online to an IOT Gecko server. To produce the desired result, the IOT gecko server posts this data online.

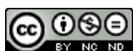

The inability to purposefully and independently operate your muscles is known as immobility. It may be either transitory or ongoing. The most frequent causes are complicated diseases, spinal cord damage, and stroke. Paresis, a severe disability, is a condition in which all mobility is lost completely. Most frequently, disruption to the neurological system, particularly the spinal cord, results in paralysis [7-9].

The nervous system is damaged or afflicted with a condition that results in paralysis, which implies that the nerve impulses going to the limbs are disrupted. Therapy aims to assist a person in adjusting to life with paralysis by keeping individuals as autonomous as practicable, even though there are cutting-edge methods for healing or managing paralysis patients. We need help with the size and cost of the machinery currently built for these gadgets. They appear restricted to medical usage and are not usable at the care facility or at their convenience. Our objective is to create a gadget that can retrain a patient's mobility while allowing individuals to use it independently and keeping the cost low sufficiently so they would pay for it out of pocket [10].

This technology also handles the circumstance in which no one is available to assist the patient, delivering a message over GSM of what he wishes to say via SMS. Our suggested system operates by detecting that the user part is actually tilted. This device's operation is demonstrated by keeping the knuckles of the mobile arm. To communicate a message, the user only needs to tilt the gadget at a specific angle. The message is conveyed differently depending on the way the gadget is tilted. Here, the characteristics of mobility are measured using an altimeter. This information is then transmitted to the microcomputer [11-14]. The microprocessor analyses the data and presents the specific message per the input received. The corresponding information is now shown on the LCD screen by the microprocessor. As soon as the gyroscope sends a motion indication, it also emits a buzzing and a text.

The patient can opt to rotate the gadget for an additional period, which could cause an SMS to be transmitted via a Mobile phone to the authorized caregiver of the patient with the information which the patient wants to express if there is no one available to attend towards the up these issues on the LCD. In this approach, the Autonomous Paralysis Health Care System regulates the person's ability to take care of themselves, ensuring prompt attention and, in turn, the patient's overall health. Patients with paralysis can benefit from this device [50]. Whenever they require assistance, they may ask by making certain gestures. They can live in this environment like any other person using this motion detection. This technology is sturdy, lightweight, and affordable. So that they can purchase debt-free, paralyzed individuals will be able to move thanks to this device independently. Just thought this task's nature and form differ from individual to individual does not mean that it is insignificant.

As technology leaders, we are responsible for creating new technology to assist patients with paralysis. Various options are thus needed to help these patients. The microprocessor can be used to build this system shortly. The chip contains every component, so because we can. The paralyzed patient can use this chip with ease. Avoid using sleeves and wristbands. However, there seems to be one drawback that will materialize: cost increases [16].

### 1.1. Objective

Medical organizations would have been forced to reduce nursing personnel for patients due to rising labor costs. The goal of our initiative is to provide fresh innovations for use in routine nursing home care. This study presents a safe IOT-based system for tracking and facilitating paralyzed patients' healthcare. It enables us to manage clinical outcomes without a nurse. With individuals who have paralysis, there are many different types of devices available to keep their bodies functioning normally instead of conversing.

For example, "Acurpo Care System ACS Entryway Gym Rope Workout for Elbow" and "Circular Palm and Thumb Workout to Improve Fingers." Numerous more kinds of equipment are also available for paraplegic patients undergoing physiotherapy. Physicians and nurses will be present to communicate with patients and recognize their requirements and emergencies, not technology or systems. In the event that a victim has a need when in a crisis, a physician cannot always be with them; otherwise, the scenario becomes dangerous.

Therefore, various approaches are required to support these individuals, and it is our responsibility as aspiring engineers to create new technology to assist those who are paralyzed. Thus, individuals with paralytic in Grades A and B will benefit significantly from this gear, which combines both hardware and software. It will be constructive for communicating both emergency cases and fundamental needs. It is also affordable to buy and simple to manage. IoT-based technology can be used by both educated and uneducated folks.

The suggested technique enables basic hand movements for paralyzed patients to communicate. Each sensor is connected to a certain finger because of the way the inertial sensors are positioned on the gloves. These accelerometers are attached to the Atmega 32B-powered Arduino Board with the aid of connecting cables. When the accelerometer's orientation is altered, the baseline or constant reading of the sensor alters. The pre-coded phrases, like "contact the physician" or "critical," are shown following this value. The buzzer triggers the alarm when the text is shown to inform the patient's carers.

### 1.2. Scope

One of the more frequent problems brought on by a stroke is paralysis or the incapacity of a limb to move. Shortly after a stroke, up to 9 out of 10 patients with stroke experience some level of paralysis. Even years after a brain, stroke patients can recover spontaneous mobility with ongoing rehabilitation and treatment. Our concept is made with the purpose of someone who has had a mild hemorrhage or nerve damage. We've created a system that allows someone who has had a cerebrovascular disease or partial paralysis to converse with someone in case of emergencies by simply moving his head. This system lets the individual interact with someone without needing assistance with basic tasks like having turned on the illumination or adapting the bed. He need not even communicate to request quick assistance.

Additionally, we will gather real-time data on the patient's condition metrics and deliver an alert signal to the victim's family. Since the doctor can easily watch the patient's health progress over time, this knowledge can be constructive for the practitioner in determining any assumptions and providing the patient with the appropriate medical assistance. One of the main causes of illness and fatality in adults, hemorrhage results in 17.3 million yearly fatalities. Upwards of 1.56 billion dementia patients in India are predicted to pass away as a result of something like a stroke by the year 2030. A real emergency is a stroke. Therefore, as proposed in our conceptual model, the stroke physician's patient's health evaluation, tracking, and quick responsiveness to his demands will assist in shortening the time it takes for a healthcare carer to arrive and thereby lower the death rate.

### 1.3. Problem Description

According to the vast majority of research, hospitalized patients' top requirements are self-assurance, connection, knowledge, learning, assistance for their healthcare, and soul. Also, urgency is crucial when it concerns a patient's basic needs, such as drinks, nutrition, and bathroom access. You cannot operate perhaps partly or totally the immobilized portions of the body. Physical inactivity may be characterized by a loss of consciousness depending on how the impairment happened. Catastrophes bring on temporary immobility. Furthermore, the biggest issue people have is that even if their bodies work on the inside, their knee and body motions are ineffective in communicating the needs of patients. However, one advantage is that they make a tiny hand motion that allows them to communicate their wants.

## 2. Related Work

Neither the nervous system nor persistent disability can heal on their own. Bell's palsy is a temporary impairment that typically goes away by itself. Orthopedic, cognitive, and cognitive therapies can offer remedies and assistive devices to alleviate immobilization as well as enhance recovery. Effective rehabilitation methods can enhance life satisfaction and enable people with all types of immobility to maintain their independence. The need for increased potential will depend upon the kind and severity of the condition. The physician might advise on rehab in adding up to:

1. Technology that seems adaptable and allows you to function or feed independently.
2. Wearable technology includes crutches, walkers, motorcycles, and batons.
3. Orthotic and prosthetics tools, including braces.
4. Vocal style pc, illumination, and communications technology.

M. M. Khan et al. [17] concentrated on developing and implementing an IoT-based health surveillance system. Users can choose their health criteria using an Internet - of - things gadget, which may assist them in maintaining their well-being over time. The patients might ultimately seek medical help if they are in need. Individuals could quickly and conveniently communicate the physician's medical factor data through a single application. Any physician can keep tabs on a patient's condition from the range. Their device will take a person's temp, pulse rate, and oxygen saturation levels before transmitting the information to an app over Bluetooth. The screen also receives this data, giving the individual a fast view of their present condition. With the aid of the method, older patients, those with asthmatic, Emphysema, chronic illness patients, COVID-19 patients, and those with diabetes will be capable of maintaining their long-term health.

By adopting Patient Health Tracking, which uses monitoring and the Internet to interact with loved ones in case of issues, S. R. Krishnan et al. [51] offered a creative initiative to avoid such abrupt death rates. Their system includes temperatures and pulse sensors. A microprocessor has been linked to a Display screen to monitor the health diagnosis, and a wireless connection transmits the data to a web-based server. [34]If the person's body temperature or pulse suddenly changes, IoT is utilized to notify the person. Additionally, this device transmits the Web live patient temperatures and pulse rate using date stamps.

S. M. Hadis et al. [19] created a patient tracking system that can electronically show the findings using Android applications and recognize the threshold of physiological parameters, assess the degree of physiological parameters based on the condition of the patient, as well as provide alerts for faulty conditions. This initiative would decrease the workload for nurses working and offer a far more practical way to check each participant's vital signs throughout the ward. The traditional system, which calls for a physician to visit each patient to check their heart rhythm, takes much time. With this method, nurses can keep an eye on individuals' conditions using Android applications that can be downloaded to any Android smartphone. In addition, by retrieving the information from the cloud in the shape of an excel spreadsheet, physicians or nurses can easily evaluate the prior vital sign status.

Richa et al. [20] proposed a patient medical surveillance system that can be utilized widely in real emergencies because it allows for regular inspection, recording, and database storage. Furthermore, the IoT gadget can be combined with laptop computers to transfer the database among critical care and therapy facilities. Additionally, this health is highly helpful in pandemic situations.

The most crucial measures needed for trauma patients include body temp, heartbeat, and blood oxygen. M. M. Khan et al. [21] propose an IoT-based approach that acts as a significant health surveillance system using the key parameters of patient body temp, heartbeat, and oximetry. This device contains an LCD that can be readily synchronized with a smartphone app to provide rapid access to the observed temp, heartbeat, and blood oxygen level. The suggested IoT-based technique made use of an Uno-based system and was examined and approved by five people.

S. Abdulmalek et al. [22] presented a review study that uses the IoTs to examine current trends in health - care surveillance systems. Regarding their importance and the advantages of IoT healthcare, the paper examines the advantages of IoT-based health services. Through a literature analysis, they provide a comprehensive evaluation of recent research on Internet - of - things medical surveillance systems. The literature evaluation contrasts the efficacy, effectiveness, data security, confidentiality, privacy, and surveillance of different systems. They also investigate IoT surveillance systems utilizing wireless and wearable sensors and offer a taxonomy of healthcare monitoring sensors.

A. Rohith et al. [23] built a Patients Healthcare Monitoring System by utilizing an ESP8266 and an Uno. Thing Speak was indeed the Interactive system employed for this project. Using the HTTP protocol, the IoT application and API Thing Speak store and retrieves data from connected gadgets through a LAN or the Internet. The pulse rate and temp may be monitored using this IoTs gadget. It sends data to an IoT platform while continuously tracking the air temperatures and pulse rate. The oximeter detector can detect Pulse Beat (BPM) and Elevated Heart (HR/BPM). The LM35 sensor module is used to calculate body temperature. The patient must be housed in a space maintained at a particular humidity level and temperature ratio. As a result, the patient does not feel uncomfortable in the space.

In some circumstances, assistive technology may have mobility issues. In some situations, manual bicycles can provide excellent mobility. However, size restrictions for wheelchairs might not be adequate for you. They may harm your spine, irritate your skin, and lead to ulceration in certain users. Limited supply and a wide range of options. Braces and other orthotic devices: These factors contribute to pain in the entire limb, lumbar pain, poor equilibrium stability, or dread of falling, general exhaustion and functional decline, aggravation, as well as skin problems, soreness, or discomfort.

Voice recognition software for phones, lights, and computing your phrases will only sometimes display perfectly on the monitor. It's possible that voice recognition would not be capable of picking out the correct phrase. For example, it occasionally finds it difficult to differentiate between the synonyms "there" and "their." Furthermore, it may experience problems with acronyms, jargon, and technical terms.

## 3. Proposed Methodology

In this research, individuals are often unable to communicate their demands because they are unable to talk clearly or use multiple languages due to a loss of muscle function in their brains. In such a case, we suggest a system that enables a disabled person to move any portion of his body capable of moving to broadcast a statement on the LCD. This method also addresses the circumstance in which nobody is available to take care of the individual and transmits a message via GSM, a digital cellular operator popular among smartphone users, so the sick might express his needs via SMS. Furthermore, we may continuously watch the paralyzed patient's movements through the website. Here, an RF module is used to transport data from the sender to the recipient, while an Arduino microcontroller handles a single device function, as displayed in Fig.1.

We used a MEMs sensor to recognize mobility in the transmitter kit. Through all the Rf transmitters, the receiving kit would receive this data. The patient's side is where this kit is available. We will use the Gsm modem in the receiving kit to transfer the information to the website [5, 24-27]. Furthermore, this device is less expensive than current apparatus like exercise and analysis tools. The suggested technique enables basic hand movements for paralyzed patients to communicate. Each sensor is connected to a particular finger because of how the inertial sensors are positioned on the gloves. With the aid of connecting wires, this sensor is attached to the Atmega8-powered Arduino UNO. When the accelerometer's direction changes, the initial or steady measurement of the sensor changes [15, 28-33]. This value determines which well before warnings, including such as "contact the physician" or "urgent," are presented. We suggest a method that enables a disabled individual to move any motion-capable component of his body to easily convey text across the screen.

The proposed system also handles the circumstance in which no one is available to care for the patient, delivering a message over GSM of what he wishes to say via SMS. Our suggested system detects whether the user part is tilting in a direction. The device's operation is demonstrated by grasping the gadget at the fingertips of the moving hand. The user generally wants to position the device at a precise angle to send a text message. The message is conveyed differently depending on the way the gadget is tilted [42]. Here, the characteristics of motion are measured using an accelerometer. This information is then transmitted to the microprocessor. According to forecasts, the newly suggested system will result in the following:

1. It includes a Wi-Fi module that enables the remote gathering of information and creation by allowing communication with sensors put on the board through Wi-Fi.
2. It is affordable so that the wealthy and the poor can use it if their needs are known.
3. It is an alphanumerical device used by both educated and uneducated folks.
4. It is a palm gadget that may be readily transported anywhere.
5. Despite the fact that some of the patients' body parts may not be functioning, they may still listen and comprehend how the process works.

This study uses two block diagrams, one of which shows the blocks that make up the arm of the doctor, while the other of which shows the arm of the patient. The voltages regulator, inverters, and circuits were used to construct the power distribution circuits. A lasting dc voltage is obtained by starting with an air conditioner voltage, adjusting it, detaching to a dc level, and then successfully obtaining a desired settling dc voltage. The orientation is often obtained via a Board voltage control board, which accepts a dc voltage and outputs a relatively lower dc voltage. This voltage remains constant despite changes to the information dc voltage or the yielding stacking connected with the voltage source.

A device called a transformer converts live electrical AC into reduced voltage AC or the opposite way over. Most likely, we'll switch from live electrical AC to low-power DC. Therefore, there is unquestionably no reason to use an advanced transformer [35-39]. The voyage voltage down typically travels down the information Voltage controlled, which is used as a component of the energy supply. The ratio of the turn's primary and secondary windings determines how far the distributor lowers the energy. Observe the oscillating flag's magnificence well before the converter component. When compared to the flag just after the converter component chart, their size is relatively large. This demonstrates that the flag was flown all around the generator. It becomes obvious that there is a question about why a transformer is used in this system. The following are the main justifications for using a converter in the system. The power supply that we receive from the Ac source must be reduced. Transformers are capable of lowering voltage levels straightforwardly and efficiently. The converter square's diodes cannot withstand the abnormally high voltage coming from the AC mains. In this manner, the converter is first circumvented by the power, and the rectification area is then attached to the voltage level. The specifications of the transformer are summarized in Table 1.

**Table 1. The specifications of the transformer**

| S. No | Factor | Range |
|-------|--------|-------|
| 1 | Frequency | 50Hz |
| 2 | Rated power | 24VA |
| 3 | Input voltage | 230V |
| 4 | Output voltage | 12 V |

An LCD screen is an electronic display with a wide range of applications. It is an important module used in diverse circuits and gadgets [37,40,41,43,44]. Such modules are superior to conventional LEDs of seven parts and other parts. Depending on the ATmega328P, the Arduino Microcontroller is a compact, comprehensive, and breadboard-friendly gadget introduced in 2008. It offers the same interfaces and specs as the Arduino microcontroller in a more convenient format.

Fig.2. represents the flow of IoT-Based Solutions for Paraplegic sufferers. The Arduino has 30 male I/O connectors arranged in a DIP-30-like layout and configured utilizing Arduino software. It is actually shared by every device and is accessible either on or offline. The circuit can be supplied by a 9 V battery or a type-B micro connection. Establishing connectivity with the other actuators and pcs is possible using the Nano Board. The electronic pins P.I.N. 0(Rx) and Vcc pin is used for the serial port (Tx).

Where Rx is being used during data reception and Tx is responsible for data transfer, The Arduino IDE now has a serial monitor that can send and receive text messages from and to the board. The software also comes with FTDI drivers, which function as a simulated serial port as data is exchanged between an FTDI and USB connection to the laptop, and a Light on the Tx and Rx pins flashes. To communicate serially between both the device and the pc, the Arduino Software Parallel API is used. Network Card, widely known as a serial to Wireless component, is a piece of Iot's data link. The purpose is to transform a serial port into an embedded module that can comply with the Wi-Fi wireless transmission medium and has built-in TCP/IP and 802.11a B.G.N. network protocols.

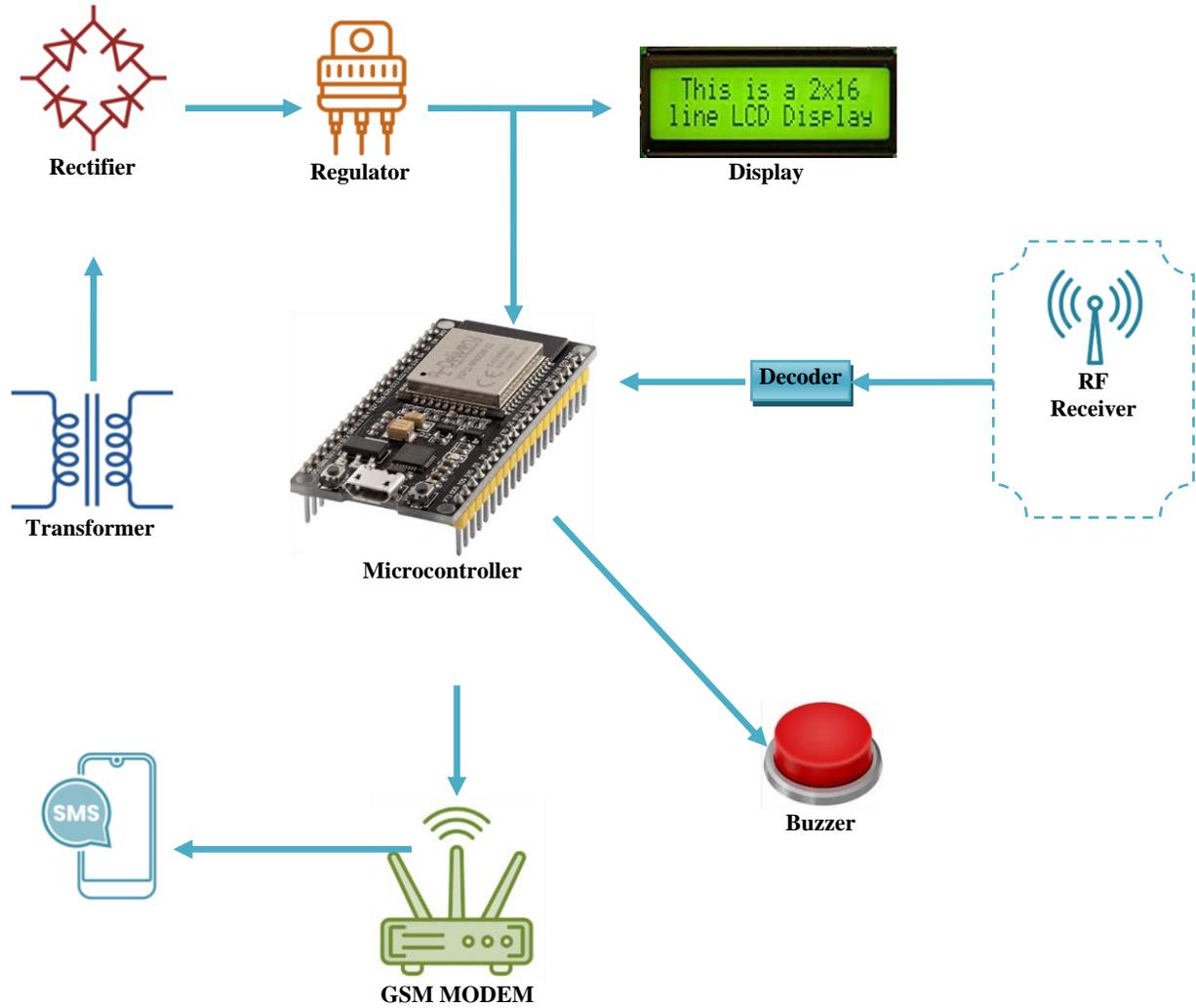

**Fig. 1 Architecture of the Proposed Model**

The detectors used in MEMS, a chip-based innovation, consist of a hanging mass between two reactive plates. This dangling material causes a voltage difference when the detector is tilted. Next, a variation in inductance is used to quantify the difference that was formed.

## 4. Result

The IOT-based paralyzed patient health care system is a software program created to assist the patient in communicating with physicians, nurses, or family members at work or home over the web. To accomplish this capability, the system uses electronics centered on a microprocessor. It uses a listener plus broadcaster circuit and a hand gesture recognition circuit. The hand gesture circuit uses a sensor and gyro to identify arm movements and sends this data wirelessly via RF to the transceiver.

The receiver system is intended to take these instructions, interpret them, show the outcomes on the LCD screen and communicate information online to an IOT Gecko gateway. The IOT gecko host subsequently posts this data online to achieve the intended result [45-48,49]. The device's operation is demonstrated by grasping the device at the fingertips of the moving hand. To communicate a message, the patient needs to tilt the gadget. Several messages are sent when the gadget is tilted at multiple angles.

Figures 3 to 7 represent the working model and the messages sent when the gadget is tilted at multiple angles. Here, we analyze the mobility parameters using an accelerometer. This information is then transmitted to the microcontroller.

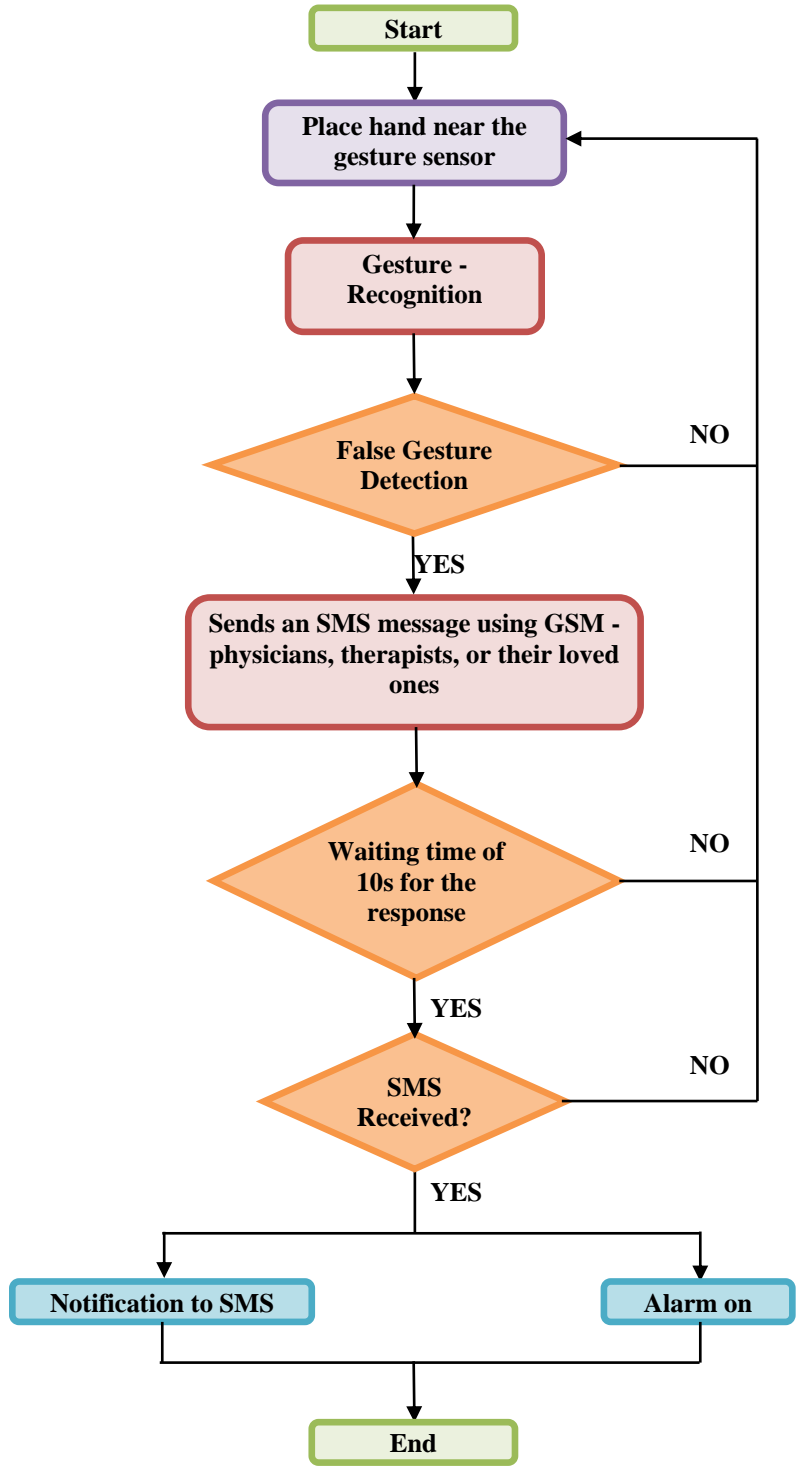

**Fig. 2 Flow of IoT-Based Solution for Paraplegic Sufferer**

The microprocessor analyzes the information and displays the specific message according to the input received. The appropriate message is now shown on the LCD by the microprocessor. When it receives a mobility signal from the sensor, it also emits a buzzer-like sound coupled with either text. The person can opt to tilt the gadget for an additional period, which will cause an SMS to be sent using GSM technology to the authorized caregiver of the patient with information that the patient wishes to convey if there is no one to listen to the up these issues on the screen.

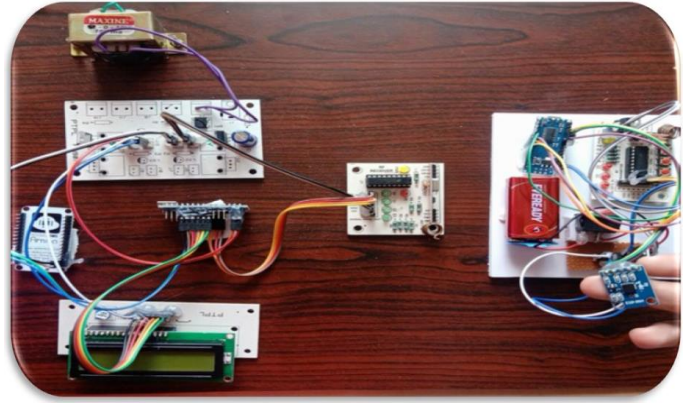

**Fig. 3 Working Model**

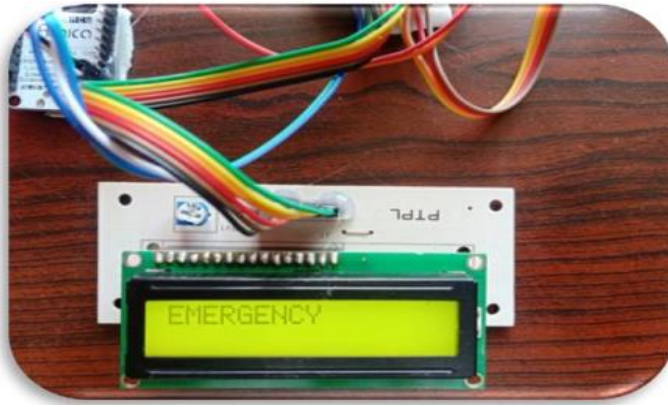

**Fig. 4 Communicates Message as Emergency**

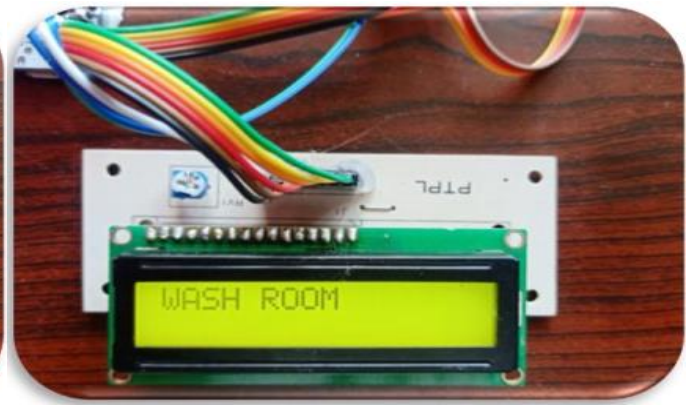

**Fig. 5 Communicates Message as Wash Room**

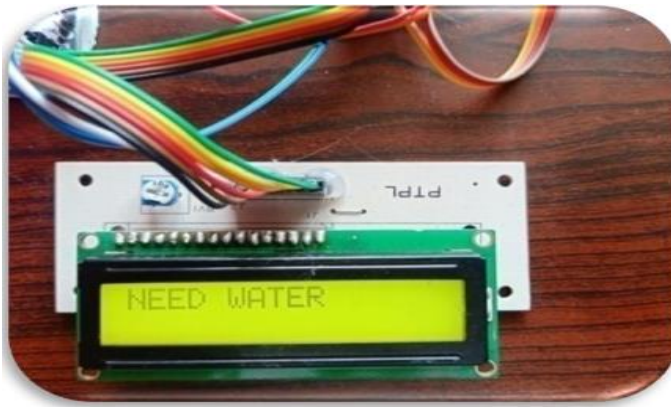

**Fig. 6 Communicates Message as Need Water**

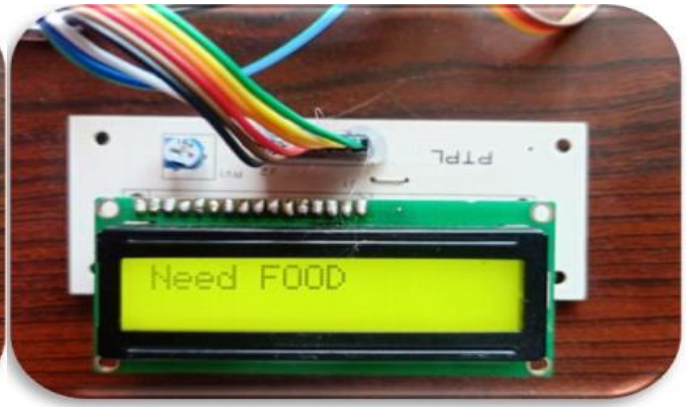

**Fig. 7 Communicates Message as Need Food**

The transistor initially provides power, which travels to the electrical supply. The Transmitter then communicates the message, which travels to the transceiver, as displayed in Fig. 8. Finally, the message travels to the Digital display, where we can view the person's basic needs. Additionally, thanks to the Wireless module, we can view the person's basic needs inside the phone service by activating the Wi-Fi connection and replicating the Port number in Browsers. Where a wireless transmission device, or Wi-Fi modules, is used, as displayed in Fig. 9. Additionally, a siren is installed so those nearby would notice and understand the requirement. This article is all about how the Paralysis Patient Basic Requirements System functions.

The incapacity to purposefully and independently operate the arms is known as paralysis. It may be either transitory or ongoing. The most frequent causes are complicated diseases, spinal cord damage, and hemorrhage. Paresis, a severe paralysis, is a condition in which all mobility is lost completely.

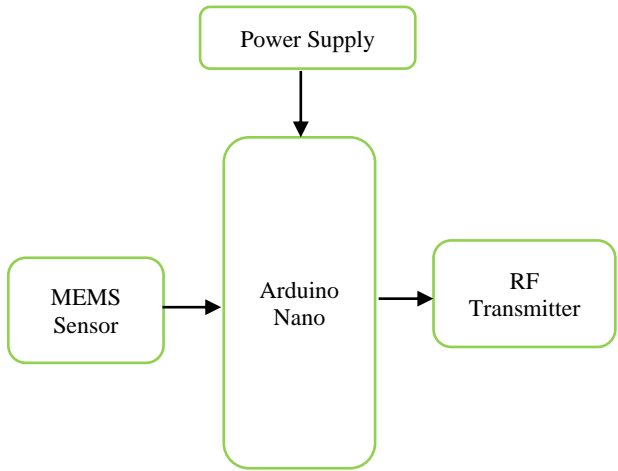

**Fig. 8 RF Transmitter**

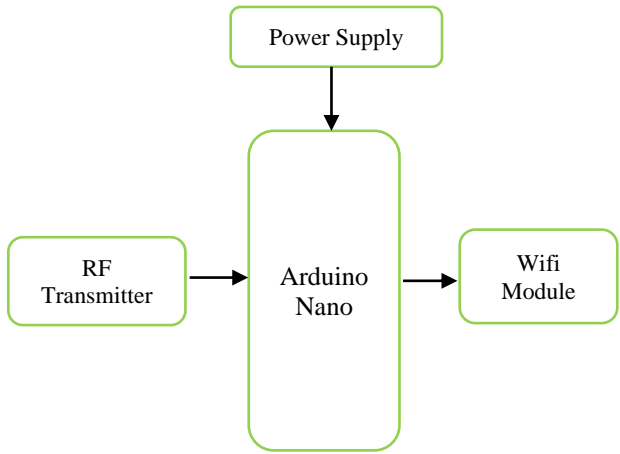

**Fig. 9 RF Receiver**

Most frequently, disruption to the neurological system, particularly the nervous system, results in immobility. The central nervous system, which is injured or ill, results in polio, which implies that the nerve impulses going to the legs are, disrupted [6]. Therapy aims to assist a person in adjusting to life despite immobility by keeping them as autonomous as practicable, even through cutting-edge methods for curing or managing polio patients. We need help with the size and cost of the equipment currently built for this kind of technology. They are restricted to medical usage and are not usable at the client's home or at their convenience. The objective is to create a gadget that can help individuals retrain their movements while also enabling them to operate it independently and making it affordable enough for individuals to do so without incurring significant debt.

## 5. Conclusion

Patients with paralysis can achieve physical autonomy thanks to this system. Whenever they require assistance, they can ask by making specific movements. Using this motion tracking, they could also live in this environment like any other person—patients whose entire or a portion of their physique has been crippled by a polio assault. Due to a lack of motor coordination by their brains, these persons are typically unable to communicate their requirements because they are either capable of speaking clearly or using hand signals. In this case, we suggest a system that enables a disabled person to show a statement on a Display screen with just a single move of any portion of his anatomy capable of gesture. This system also handles the circumstance in which no one is available to care for the patient, delivering a message over GSM of what he wishes to say via SMS. The device's operation is demonstrated by grasping the gadget in the knuckles of the moving hand. The user must bend the gadget at a specific angle to communicate a message. The message is transmitted differently depending on the way the gadget is tilted.

These folks require a variety of support, and it is our responsibility as aspiring scientists to create innovative solutions to assist paralyzed patients. Individuals with a disability can really benefit from this device. When individuals require assistance, individuals may request it by making certain gestures. By using this motion control, individuals could also live inside this environment like any other person. This technology is sturdy, lightweight, and affordable. So they can purchase debt-free. Chronic immobility cannot be cured. The brain stem cannot recover on its own. Bell's palsy is an example of a temporary disability that frequently resolves over age. Therapies, adaptive equipment, and orthotic technologies can be provided by mental, occupational, and linguistic therapists to adapt to immobility and put in more effort. The microprocessor can be used to build this system later on. The kit includes every component so which we can. The paralyzed person can use this microchip with ease. Bracelets and elbow pads should not be worn.